\documentclass[amsmath, amssymb, amsfonts, aps, prl, showpacs, twocolumn, reprint, floatfix,superscriptaddress]{revtex4-2}
\usepackage[pdftex,plainpages=false,colorlinks=true,linkcolor=blue, citecolor=blue, urlcolor=blue]{hyperref}

\usepackage{amssymb}
\usepackage{epsfig}
\usepackage{graphicx}
\usepackage{amsmath}
\usepackage{array,color}
\usepackage{natbib}

\usepackage{footmisc}

\newcommand{\im}{\mathrm{Im}}

\begin{document}

\title{Inexorable Edge Kondo Breakdown in Topological Kondo Insulators}

\author{Zhihui Luo}
\affiliation{School of Physics, Sun Yat-sen University, Guangzhou, Guangdong Province 510275, China
}

\author{Michel Ferrero}
\affiliation{CPHT, CNRS, Ecole Polytechnique, IP Paris, 91128 Palaiseau, France}
\affiliation{Coll\`ege de France, 11 place Marcelin  Berthelot, 75005 Paris,  France}

\author{Dao-Xin Yao}
\email[Corresponding author: ]{yaodaox@mail.sysu.edu.cn}
\affiliation{School of Physics, Sun Yat-sen University, Guangzhou, Guangdong Province 510275, China
}

\author{Wei Wu}
\email[Corresponding author: ]{wuwei69@mail.sysu.edu.cn}
\affiliation{School of Physics, Sun Yat-sen University, Guangzhou, Guangdong Province 510275, China
}
\affiliation{CPHT, CNRS, Ecole Polytechnique, IP Paris, 91128 Palaiseau, France}

\date{\today}

\begin{abstract}
Kondo breakdown is one of the most intriguing problems in strongly correlated electron systems, as it is rooted in many anomalous electron behaviors found in heavy-fermion materials.
In Kondo lattice systems,  Kondo breakdown can arise from either strong magnetic frustrations or critical fluctuations of collective modes. Here, we reveal a new type of Kondo breakdown  with a fully different origin in interacting topological Kondo insulators. By employing numerically exact quantum Monte Carlo simulations, we show that with open boundary conditions, Kondo screening is inexorably destroyed by interaction effects on edges or corners in these systems.  We argue that the Kondo breakdown is enforced by the symmetries of the system, because the ground states are symmetry protected Haldane phases.
\end{abstract}


\maketitle

\paragraph{Introduction-}
Kondo screening involves scattering between localized magnetic moments and conduction electrons. At low temperatures,
local moments can sometimes be completely quenched by screening and turned into spinless scattering centers.
Kondo systems can then be described by Fermi liquid theory and have a non-degenerated singlet ground state~\cite{nozieres,PColeman2007review}.
Heavy fermion metals,
which  usually consist of a lattice of $f-$ spins immersed in a sea of  itinerant conduction electrons,
 are a remarkable manifestation of such a picture.
In heavy fermion systems~\cite{stewart1984}, a highly non-trivial and challenging question is how  the screening between $f-$ and conduction  ($c-$)  electrons can be broken and the standard
Landau Fermi liquid theory no longer apply~\cite{stewart2006,coleman2019}.

There are two widely accepted scenarios for Kondo breakdown (KB) transitions in Kondo lattice models (KLM), or the periodic Anderson models (PAM) so far. Senthil and co-workers~\cite{senthil,vojta2010} proposed that
in the presence of strong quantum fluctuations, Kondo screening can be destroyed in the manner of an orbital selective Mott transition (OSMT)~\cite{pepin,deleo2008}.
The essential idea is that in an $f- c$ two-orbital lattice system with
sufficiently strong fluctuations, the  $f-$ orbital can acquire a diverging self-energy at low-energies as it Mott localizes. Below a certain energy scale, this diverging self-energy renormalizes the $f-c$ hybridization, or equivalently, the Kondo screening
to zero. In this scenario of KB, the geometrical frustration usually plays a vital role, since it can greatly amplify magnetic fluctuations.
The $f-$ moments decoupled from the $c-$ electrons in a KB phase can consequently form exotic fractionalized spin-liquid states~\cite{hofmann2019} due to frustration.

Another scenario for KB is put forward by Qimiao Si and co-workers~\cite{si2001locally,si2010heavy}, which suggests that the Kondo screening can be destroyed at an antiferromagnetic (AFM) phase transition by critical spin fluctuations.
A characteristic feature of this scenario is that the two transitions, KB and AFM ordering, must coincide in a finite regime of  parameter space. Si \textit{et al}  have shown that in the extended dynamical mean-field theory (EDMFT)~\cite{antoine1996}, KLM near an
AFM phase transition
can be mapped into a microscopic sub-ohmic Bose-Fermi Kondo model which has a Kondo breakdown fixed point~\cite{Si2003, Pankov2002, zhu2004}. In particular, for the temperature-dependent spin susceptibility $\chi(T)^{-1} = \Theta +AT^{\alpha}$, EDMFT finds an anomalous exponent $\alpha\approx 0.75$, which is in remarkable agreement with the experimental value $\alpha = 0.75 \pm 0.05 $ found in the $\mathrm{CeCu_{5.9}Au_{0.1}}$ material~\citep{Schroder1998, schroder2000onset}.
In addition, in  $\mathrm{CeCu_{6-x}Au_{x}}$, $\mathrm{CeRhIn_5}$~\cite{shishido2005drastic}, or doped $\mathrm{YbRh_2Si_2}$~\cite{si2010heavy}, the sudden collapse of Fermi surface that signals a KB,  is indeed observed in the vicinity of the onset of AFM order. It is notable that similar to the spin fluctuations, the critical charge fluctuations may also give rise to KB~\cite{coleman2019}.
From a unified perspective, the above two scenarios can be summarized as a global phase diagram in the parameter space of degree of quantum fluctuations $G$, which represents magnetic frustration or spatial dimensionality, and the Kondo coupling strength $J_k$~\cite{si2010review}. In this global phase diagram, the two different scenarios for KB can be understood as two different quantum transition trajectories~\cite{si2010review}.


In this work, we introduce a novel type of  KB that can occur in certain interacting topological Kondo insulators (TKI)~\cite{dzero2016}.
By using numerically exact determinant quantum Monte Carlo (DQMC) simulations~\cite{bss1981,assaad2002}, we investigate  correlation effects in two periodic Anderson models (PAM): the one-dimensional interacting topological Kondo insulator (1d-TKI)~\cite{lobos2015}, and the two-dimensional interacting quadrupole TKI (2d-QTKI)~\cite{Benalcazar61,Seshadri2019}. We show that with open boundary conditions (OBC), a Kondo singlet ground state is unstable in these systems. A KB
is inexorably happening on edge sites of the 1d-TKI model, or corner sites of the 2d-QTKI model.
We find that like the 1d-TKI model~\cite{lobos2015}, the interacting 2d-QTKI model  has a  Haldane phase ground-state.
We argue that in these models the occurrence of KB is dictated by the symmetry-protected topological (SPT) Haldane phase ground state~\cite{pollmann2012, chenxie2013}.
In this sense, edge KB is ``protected'' by system symmetries, hence insensitive to the details of system parameters such as disorder, interacting strength, etc.
The edge KB revealed here therefore represents a new mechanism for Kondo effects breakdown that is distinct from the aforementioned OSMT or critical fluctuation scenarios.

\begin{figure}[t]
\includegraphics[scale=0.38]{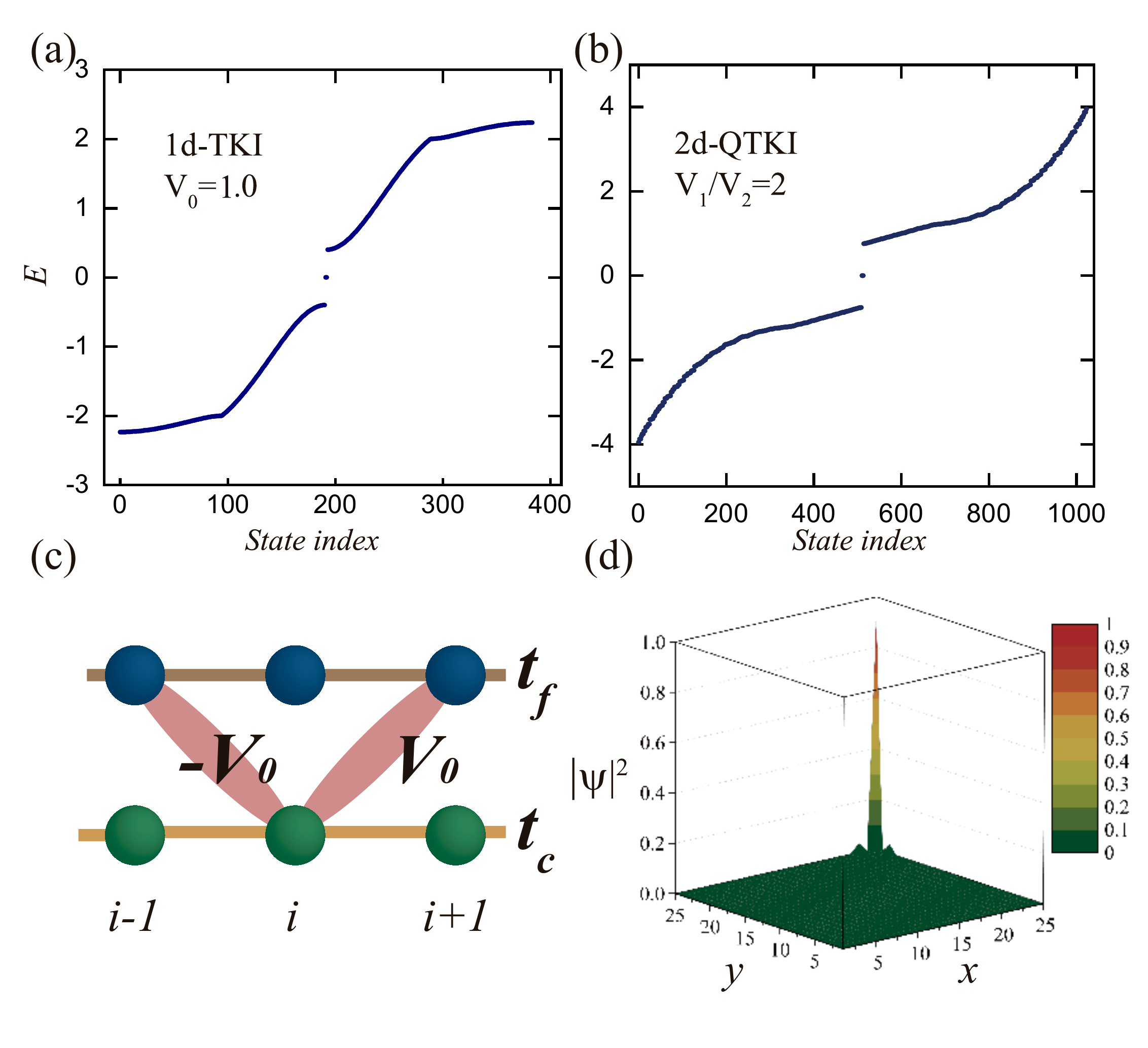}
\caption{Non-interacting properties of the 1d-TKI and 2d-QTKI models. \textbf{(a)} and  \textbf{(b)}: Energy spectra of ${\cal H}\left(k\right)_{1D}$  and ${\cal H}\left(k\right)_{2D}$  respectively show edge modes of the 1d-TKI and corner states of the 2d-QTKI with OBC. \textbf{(c)}: Illustration of the periodic Anderson model for the 1d-TKI. \textbf{(d)}: Probability amplitude of a corner eigenstate in real-space, showing the it is essentially localized at the outermost corner site.
}
\label{fig:z1d}
\end{figure}

\paragraph{Models and methods-}
We use periodic Anderson models to construct the one-dimensional (1D) and two-dimensional (2D) interacting topological Kondo insulators,
\begin{eqnarray}
H &&=H_0+H_U,  \nonumber \\
H_0&&=\sum_{k,\sigma}\Psi^{\dagger}_\sigma(k){\cal H}(k)\Psi_\sigma(k),\ H_U=U_f\sum_{i}n^{f}_{i\uparrow}n^{f}_{i\downarrow }, \nonumber\\
\Psi_{1D}&&=(f, c), \quad \Psi_{2D}=(f_{1}, f_{2}, c_{1}, c_{2}), \nonumber\\
{\cal H}\left(k\right)_{1D}&&= \left(\begin{array}{cc}
                h_f(k) & V_0(k) \\
                V_0(k) & h_c(k) \\
              \end{array} \right),  \nonumber\\
{\cal H}\left(k\right)_{2D}&&=\left(\begin{array}{cc}
                h_f(k)\sigma^z+V_2(k)\sigma^y & V_1(k)\sigma^x\\
                V_1^*(k)\sigma^x & h_c(k)\sigma^z+V_2(k)\sigma^y\\
              \end{array}
            \right).\nonumber\\ \label{eq:ham}
\end{eqnarray}
where ${\cal H}\left(k\right)_{1D}$ describes a 1D Su-Schrieffer-Heeger (SSH) like two-band TKI model~\cite{lobos2015} with $V_0(k)=2iV_0\sin (k)$, and ${\cal H}\left(k\right)_{2D}$ is a four-band second order topological insulator~\cite{Benalcazar61,Seshadri2019,lu2020}. Here the  ${V_1(k)=2V_1(\sin k_x-i\sin k_y)}$ hybridization term can be interpreted as spin-orbital coupling that is widely used in usual (first order) topological Kondo insulators, while the $V_2(k)=2V_2(\cos k_x-\cos k_y)$ term is essential to generate the quadrupole (or second order) TKI~\cite{Benalcazar61}. In other words, the $V_2(k)$ term of  \textbf{${\cal H}\left(k\right)_{2D}$} can give rise to  zero dimensional topological corner states when open boundary conditions are applied on both $x-$ and $y-$ directions of the 2d-QTKI, as shown in Fig.~\ref{fig:z1d}. For both ${\cal H}\left(k\right)_{1D}$  and ${\cal H}\left(k\right)_{2D}$, $h_f(k)=-2t_f(\cos k_x+\cos k_y)+\mu_f$  and  $h_c(k)=-2t_c(\cos k_x+\cos k_y)+\mu_c$ are respectively the  $f-$ and $c-$ intra-band dispersions. We set $t_c = 1 $ as energy unit throughout this work, and without loss of generality, $t_f$ is fixed to $t_f=0.2$ denoting a narrow $f-$ band dispersion. To study interaction effects, we use the numerically exact determinant quantum Monte Carlo (DQMC) simulation  for finite temperature~\cite{bss1981} and projector quantum Monte Carlo (PQMC) for zero temperature computations~\cite{assaad2002} . The DQMC time discretization is set to $\Delta \tau = 0.05$ and for each data point we use typically millions of Monte Carlo sweeps to reach the required precision.

\begin{figure}[t]
\includegraphics[scale=0.95]{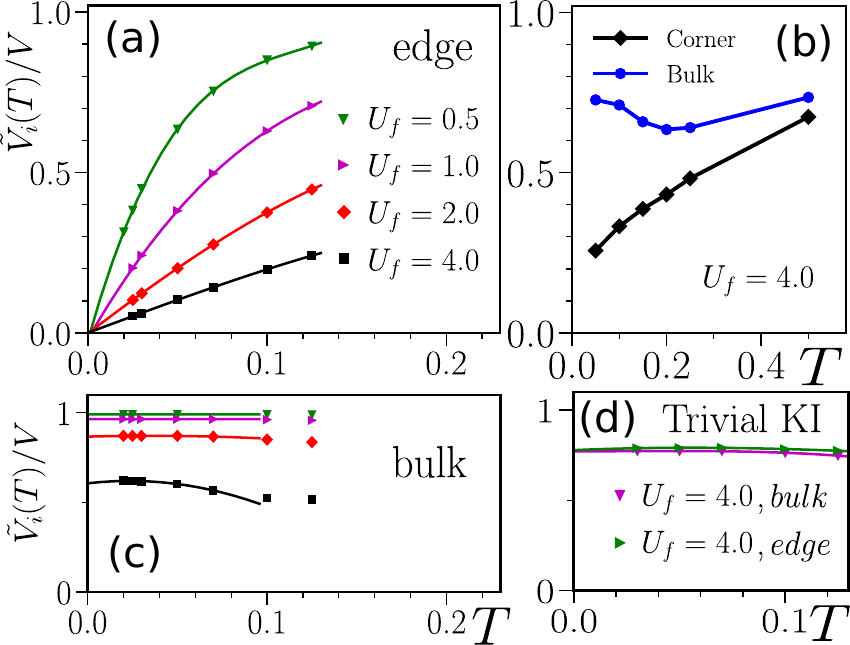}
\caption{Low energy effective $f-c$ hybridization $\tilde{V}(T)$ as a function of temperature $T$ of one dimensional TKI models.
\textbf{(a):} $\tilde{V}(T)$ at edge sites of the 1d-TKI.
\textbf{(c):} $\tilde{V}(T)$ at a bulk site of the 1d-TKI.
\textbf{(b):} Low energy effective $f-c$ hybridization $\tilde{V}_i(T)$ as a function of temperature $T$ for the 2d-QTKI with OBC.
\textbf{(d):} $\tilde{V}(T)$ for the trivial 1D Kondo insulator (see text). Lines show quadratic polynomial fittings.
}

\label{fig:z2d}
\end{figure}

\paragraph{Results: Effective $f-c$ hybridization in interacting topological Kondo Insulators-}
We first study the Kondo screening of the 1d-TKI model with ${\cal H}\left(k\right)_{1D}$ at finite temperatures, then extrapolate to $T=0$ to reveal the zero temperature behavior. The DQMC simulations are carried out on an $192\times 2$ sites chain with two open ends. The lowest temperature reached here is $T=0.02$, where the finite size effects are negligible since
 the inverse temperature $\beta=1/T = 50$ is significantly smaller than the length of the chain $L=192$.
We find that for the parameters we study ($V_0 =1, U= 0\sim 4$), the non-local self-energies are much smaller than the local ones. Hence the low-energy effective hybridization $\tilde{V}$ at the real-space site $i$ is essentially determined by the local self-energies $\Sigma_{ii}$,
\begin{equation}
\tilde{V}_i(T) = \frac{V}{\sqrt{Z_i(T)}}= \frac{V}{\sqrt{1-\frac{\im{\Sigma_{ii}(\omega_0)}}{\omega_0} |_T }}.
\label{eq:hyb_renorm}
\end{equation}
we note that $Z_i(T\rightarrow 0)$ becomes quasiparticle residue weight $Z$ in the Fermi liquid theory. Hence $\tilde{V}_i(T)$ extrapolated to $T=0$ estimates the zero temperature low-energy effective hybridization between the $f-$ and $c-$ electrons.
For the 1d-TKI model, we display $\tilde{V}_i(T)$ in Fig.~\ref{fig:z2d}a and Fig.~\ref{fig:z2d}c as a function of temperature $T$ for site $i$ located on the edge and in the bulk respectively.
One can clearly see that $\tilde{V}_i(T)$ at a bulk site (in the center of the chain, Fig.~\ref{fig:z2d}c) changes mildly with temperature at different $U_f$. For $U_f = 0.5 \sim 4$, all $\tilde{V}_i(T)$ approach finite values as $T\rightarrow 0$, indicating that Kondo screening is always intact in the bulk. In stark contrast to the bulk case, see Fig.~\ref{fig:z2d}a, edge sites (at the open end of the chain) display strong temperature dependence of $\tilde{V}_i(T)$, suggesting that interaction effects greatly renormalize Kondo screening on the edge. Importantly, for different $U_f$, all effective hybridizations $\tilde{V}_i(T)$ on the edges eventually vanish as $T\rightarrow 0$. Decreasing $U_f$  does not alter the edge KB in the  $T\rightarrow 0$ limit, which indicates that the edge KB occurs in the 1D TKI model at any finite $U_f$ .


We further study KB in the 2d-QTKI model, where DQMC simulations on a $16\times 16\times 4$ lattice under OBC have been carried out. In Fig.~\ref{fig:z2d}b, the result of $\tilde{V}_i(T)$ for a bulk and a corner site are plotted as a function of temperature $T$. Similarly to the 1d-TKI model, the $f-c$ effective hybridizations
$\tilde{V}_i(T)$ again have a fully different temperature dependence at different locations: at low temperatures, as $T$ decreases, $\tilde{V}_i(T)$ in the bulk grows while $\tilde{V}_i(T)$ on the corner decreases, signaling an edge KB in the 2d-QTKI model as in the 1d-TKI model case.

One may suspect that the edge KB is a consequence of edge (corner) sites having less neighboring sites than bulk sites: Due to the reduction of dimensionality, edge sites suffer stronger
fluctuations than bulk sites, hence the edge KB can happen while bulk Kondo screening remains intact.
To clarify this point, in Fig.~\ref{fig:z2d}d, we plot $\tilde{V}_i(T)$ for a normal 1D Kondo insulator model (1d-KI) with OBC and an $f-c$ hybridization $V_0(k) = 2V_0\cos (k)$. As one can see, at a typical  $U_f = 4$, $\tilde{V}_i(T)$ for bulk and edge sites are almost the same at different temperatures $T$ and both extrapolate to finite values at zero $T$, indicating the absence of edge KB in the 1d-KI model. Therefore one cannot
attribute the edge KB to the reduced dimensionality of the edge sites.
Instead, as we will show below, the Kondo breakdown observed here is related to the interaction effects of the topological edge modes in the TKI models. As such, it is absent in the 1d-KI model.

\paragraph{Results: Entropy and ground-state properties of the TKI models-}

In order to understand the striking edge KB at infinitesimal electron-electron interactions, we now investigate the ground-state properties of the interacting TKI models. We note that if  KB happens, even if it  only occurs at the edge or corner sites of the lattice, the system is deemed to transit to a new phase whose ground-state is distinct from that of a Kondo screened state. In particular, the dangling edge $f-$ moments that decoupled from conduction electrons in edge KB, can give rise to ground-state degeneracy. The degree of ground-state degeneracy can be directly identified by the  value of the zero temperature entropy $S(T=0)$, which can be obtained by extrapolating the temperature dependent entropy $S(T)$~\cite{binder1981monte} to zero $T$,
\begin{eqnarray}\label{eq:entropy}
  S(T)=N\ln (4)+ \frac{1}{T} E(T) - \int_{\infty}^{T}E(T^\prime)d(\frac{1}{T^\prime}).
\end{eqnarray}
where $N$ is the total number of sites and $\ln (4)$ is the entropy per site of a $spin-\frac{1}{2}$ fermionic system in the high-temperature ($T \rightarrow \infty $) limit. $E(T)$ is the total energy calculated by DQMC at temperature $T$.

\begin{figure}[tp]
\includegraphics[scale=0.35]{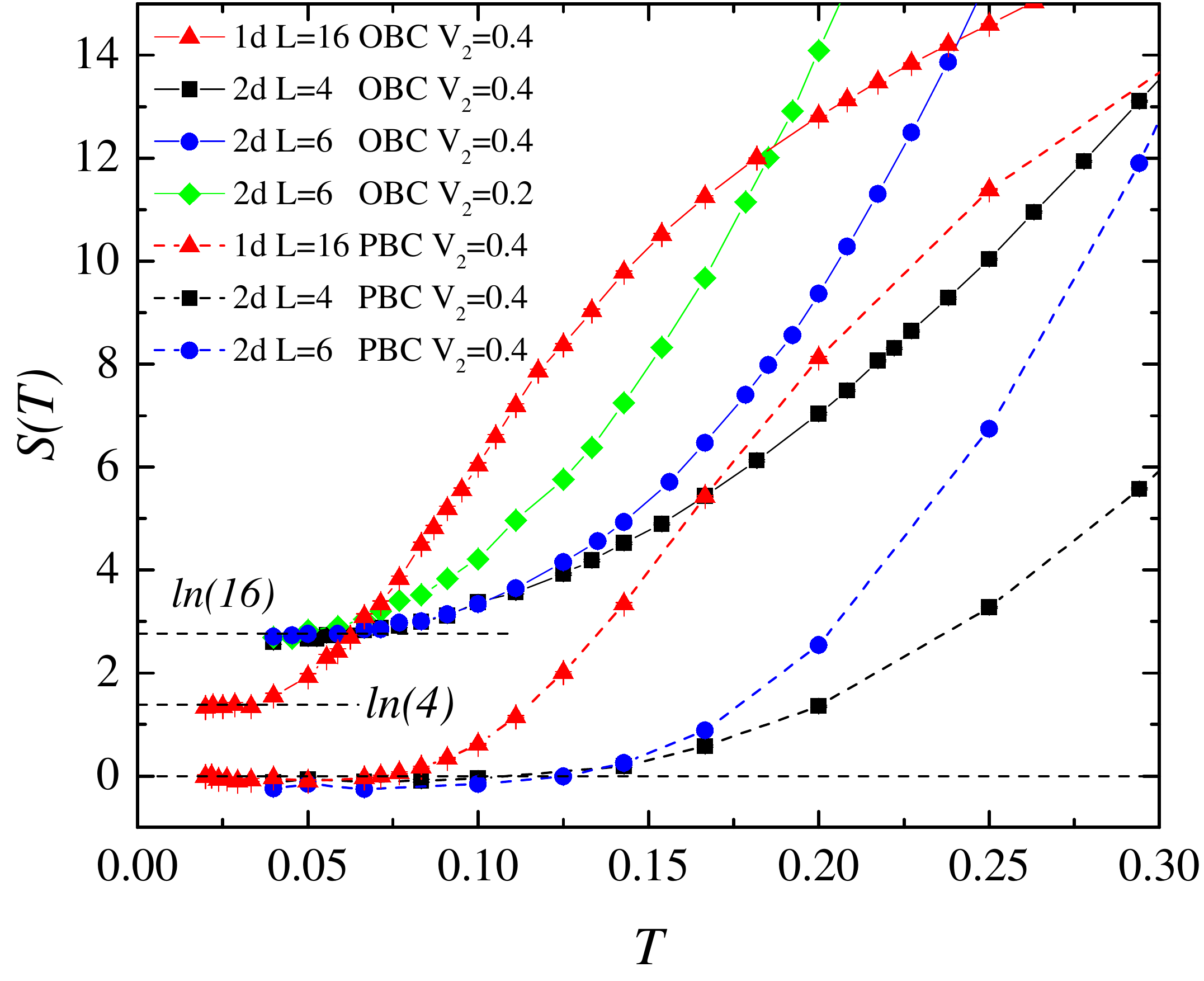}
\caption{ Entropy $S(T)$ as a function of temperature $T$. For the 2d-QTKI model with OBC, $S(T)$ approaches $\ln (16)$ as $T\rightarrow 0$, while for the 1d-TKI $S(T)$ approaches $\ln (4)$ as $T\rightarrow 0$.
For PBC, all cases (dashed lines) have $S(T) \rightarrow 0$ as $T$ is lowered. Here for OBC in the 2d-QTKI, $U_f = 4$, $V_1=0.8$, $V_2=0.4,0.2$; for PBC in the 2d-QTKI, $U_f = 4$, $V_1=0.8$, $V_2=0.2$. In the 1d-TKI case, $U_f = 4$, $V_0=0.8$.}
\label{fig:entropy}
\end{figure}

In Fig.~\ref{fig:entropy}, the $S(T)$ results from DQMC can be divided into two groups: dashed lines show interacting TKI models with periodic boundary conditions (PBC) while solid curves show cases with open boundary conditions. As we can see, $S(T)$ for PBC in the 1d-TKI and 2d-QTKI models both approach zero at low temperatures, indicating the existence of a unique ground-state in these systems. This is in agreement with our result above that KB only happens at the edge or corner sites of the TKI models. With PBC, the topological edge modes in the 1d-TKI and 2d-QTKI models are both absent, hence the systems preserve Kondo screening and their ground-states are  non-degenerate Kondo singlets.
The more compelling results for $S(T)$ are with OBC which, as a function of temperature $T$, become flat and saturate at small $T$, suggesting that the 1d-TKI model has a four-fold ground-state degeneracy [$S(T\rightarrow 0) \approx \ln(4)$], and the 2d-QTKI model has a sixteen-fold ground-state degeneracy [$S(T\rightarrow 0) \approx \ln(16)$]. Note that here the ground-state degeneracy is insensitive to the hybridization strength $V$ or to the linear size of lattice $L$. For example, for the 2d-QTKI model with $V_1=0.8, V_2=0.4$, $S(T)$ and $L=4$ (back squares) or $L=6$ (blue dots) both approach  $\ln(16)$ at almost the same onset temperature $T\sim 0.07$. Switching to a different case with smaller $V_2 = 0.2$ (green diamonds), entropy $S(T)$ again saturates at $\ln(16)$ at low $T$.

We point out that here the non-zero $S(T)$ in the $T \rightarrow 0$ limit apparently cannot be explained in the Ginzburg-Landau phase transition framework, since the small lattice size $L=4$  leaves no room  for any amplitute or  phase fluctuations to become soft.
Note that it is established that the interacting 1d-TKI model can be mapped onto a $spin-1$ Haldane chain~\cite{lobos2015}, whose ground-state is a topologically nontrivial Haldane phase protected by a set of global symmetries~\cite{gu2009,pollmann2012, chen2012}. The two unscreened $f-$ moments have long-range entanglement and the ground-state is four-fold degenerate, which agrees with our explicit computation of the entropy $S(T)$.
Here we have a similar mechanism  mapping the 1d-TKI  onto the a $spin-1$ Haldane chain: in the periodic Anderson model described by ${\cal H}\left(k\right)_{1D}$, the charge degree of freedom is frozen by the hybridization gap (see Fig.~\ref{fig:z1d}a), when $U_f$ is turned on, the spin correlations between nearest-neighboring $f-c$ sites lead to an effective local $spin-1$ state~\cite{lobos2015}. The only difference is that here the charge gap in the conduction band is embodied by hybridization while in Ref.~\cite{lobos2015} it is a dynamically generated Mott gap in the Kondo lattice model~\cite{nozieres}.
For the 2d-QTKI model with OBC, assuming four dangling $f-$ moments located at the four corners accounts for the $2^4 =16$ fold ground-state
degeneracy. Hence we conclude that the ground-state of the interacting 2d-QTKI model is in a 2D Haldane phase, resembling the two-dimensional Affleck-Kennedy-Lieb-Tasaki (AKLT) model~\cite{affleck1988,gu2009,chen2011}, in regards to the symmetry protected topological ground-state. Below we display results of magnetic correlations of 2d-QTKI to further demonstrate this is indeed true.

\begin{figure}[t]
\includegraphics[scale=0.28]{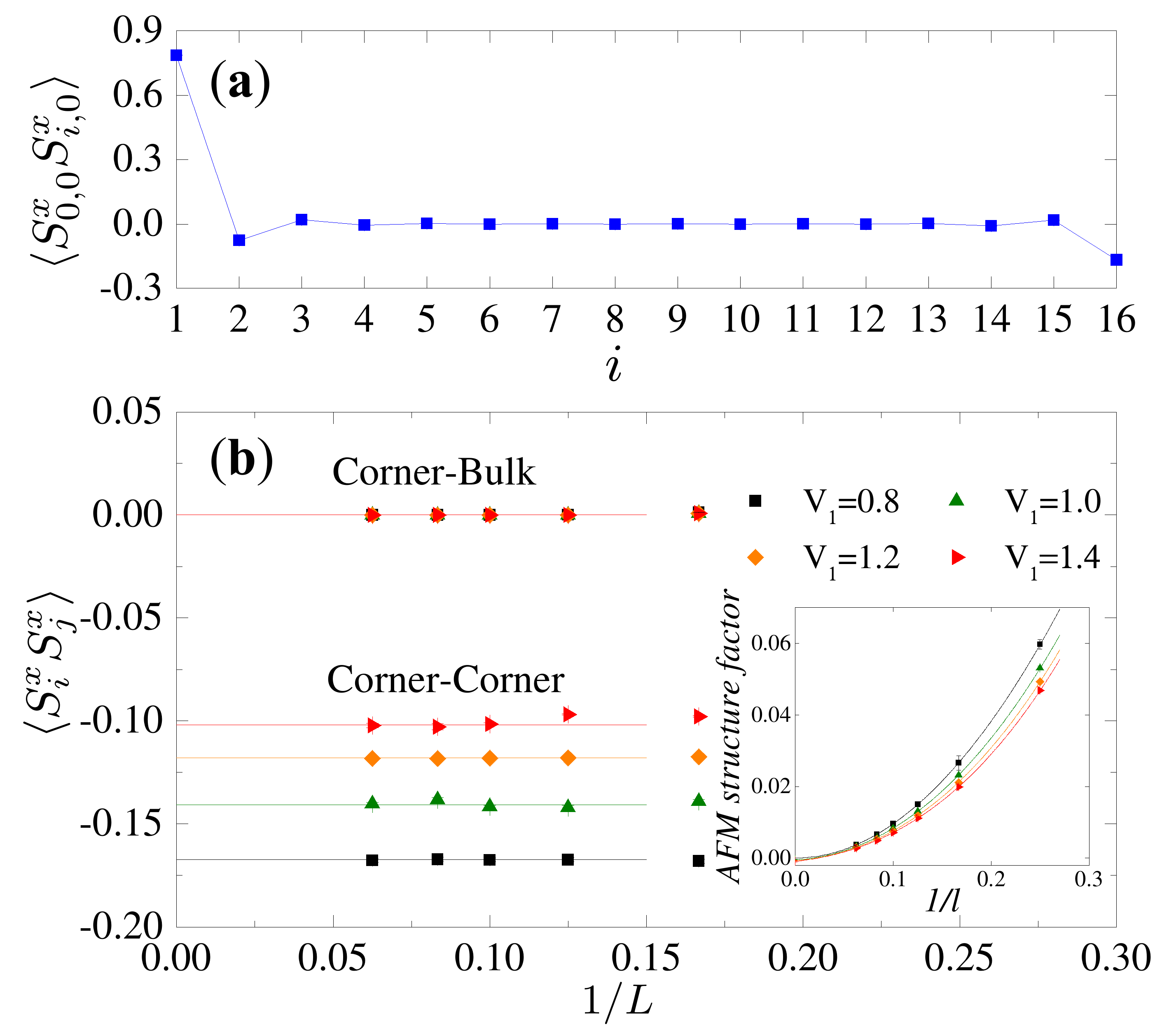}
\caption{Zero-temperature magnetic properties of the 2d-QTKI model with OBC suggest that its ground-state is a two-dimensional Haldane phase with four dangling $S=\frac{1}{2}$ spins located at the four corner sites.
\textbf{(a)} $\langle S^x_{0,0} S^x_{i,0} \rangle$  as a function of site distance $i$ between  corner site $(0,0)$ and site $(i, 0)$ on the $R_y=0$  open edge ($R_x=i$) of an $L=16$ lattice. Here we see that magnetic correlations decays rapidly with distance but are finite between two corner sites.  \textbf{(b)} Magnetic correlations between corner-bulk sites and corner-corner sites extrapolated to the thermodynamic limits $1/L \rightarrow 0$ for different values of the hybridization strength $V$, here $U_f = 4$. \textbf{Inset:} AFM structure factor extrapolates to zero as $L \rightarrow 0$ which suggest that long-range magnetic correlations between corner sites are not due to a Landau phase transition.}
\label{fig:sxsx}
\end{figure}

A characteristic feature of the Haldane phases is its long-range entanglement between the edge spins~\cite{lobos2015}.  To demonstrate that the interacting 2d-QTKI model is indeed in a Haldane phase,  we display the zero-temperature  magnetic properties of the 2d-QTKI model in Fig.~\ref{fig:sxsx} obtained by PQMC simulation. Fig.~\ref{fig:sxsx}a shows the magnetic correlations  $\langle S^x_{0,0} S^x_{i,0} \rangle$  as a function of distance $i$ between  corner site $(0,0)$ and site $(i, 0)$ on the $R_y=0$  open edge ($R_x=i$), for a typical parameter set $V_1=0.8, V_2=0.4, U_f=4$. As one can see, $\langle S^x_{0,0} S^x_{i,0} \rangle$  decays rapidly with distance $i$ and the  magnetic correlations between corner and bulk sites essentially vanishes. However, as the distance $i$ is further increased to reach another corner site ($i=16$), $\langle S^x_{0,0} S^x_{i,0} \rangle$  gains a  negative finite value, $\langle S^x_{0,0} S^x_{i,0} \rangle$  $ \approx -0.16$, showing a long-range antiferromagnetic correlation. In Fig.~\ref{fig:sxsx}b, magnetic correlations between corner-corner and  corner-center sites  are plotted at different hybridization strengths  $V_1, V_2$ with fixed $U_f =4$. Extrapolating to the thermodynamic limit $1/L \rightarrow 0 $, we indeed find a non-vanishing  long-range corner-corner magnetic correlation, while the magnetic correlation between corner-center sites is essentially zero. This result suggests that the interacting 2d-QTKI model with OBC is indeed in a Haldane phase state. We stress that the corner-center correlation vanishes in the thermodynamic limit suggesting there is no Landau long-range order in this system.  This is also evidenced in the inset of Fig.~\ref{fig:sxsx}, where the magnetic structure factor is plotted for the 2d-QTKI model with PBC, which extrapolated to zero for all the studied hybridization strength $V$.

The results above provide a roadmap to the understanding of KB in interacting topological Kondo insulators with edge/corner states:
As the repulsion between $f-$ electrons is turned on, the  $f-$ spins of the topological edge or corner modes become localized and suspended by correlation effects, which is required by the underlying Haldane phase ground state. This means that the low-energy hybridizations between $f-$ and $c-$ orbitals at edge/corner sites have to be broken to release the $f-$ spins, which in turn gives rise to a KB.  Hence, in these systems, the strong quantum fluctuations that destroy Kondo screening neither stem from magnetic frustration nor critical spin fluctuations. They are instead generated by the long-range $f-$ spin entanglement between edge or corner sites. It is worth noting that here as in the 1D spin-1 chain or 2D AKLT models, the ground-state degeneracy or equivalently the breakdown of the Kondo singlet ground-state, can in principle be stabilized by any of the $\mathbb{Z}_2 \times \mathbb{Z}_2$,  time reversal, or inversion symmetries of the systems~\cite{pollmann2012,gu2009}.

A few discussions are in order. Recently, the discovery of unusual surface transport properties of Kondo insulator samarium hexaboride  $\mathrm{SmB_6}$ ~\cite{wolgast,kim2013surface, erten2016, chen2018} has
stimulated renewed interest in Kondo breakdown problems. Although a direct Monte Carlo simulation with the full consideration of $\mathrm{SmB_6}$ lattice structure  would be difficult,  the mean-field theory studies on simplify models have been carried out
~\cite{jiang2013observation,li2014two,held2021,min2017}. It is proposed that the surprisingly  light
surface quasiparticles of $\mathrm{SmB_6}$  is due to a surface Kondo breakdown that can be attributed to
the reduction of coordination number  at the surface ~\cite{alexandrov2015,thomson2016,zhong2017}.  In this sense, the possible surface KB
of $\mathrm{SmB_6}$  is unrelated to the symmetry-enforced KB we find here. It can rather be  seen  as  a special form of OSMT  Kondo breakdown.
On the experimental side, it is predicted that higher-order topological insulators can be realized in bismuth~\cite{Schindler2, noguchi2020higher} and strained SnTe ~\cite{schindler2018}.
A similar realization of higher-order topological Kondo insulators in real materials can be challenging. However, an ultracold-atom experiment realization of
a higher-order topological Kondo insulator might be feasible~\cite{chen2016,zhang2019}.

\paragraph{Conclusions-}
We have studied the interaction effects on Kondo screening in a one-dimensional topological Kondo insulator and a two-dimensional quadrupole topological Kondo insulator. We found that the  Kondo screening can be destroyed on edge or corner sites in these systems and is enforced by the system symmetries. This special type of Kondo breakdown represents a novel mechanism for the destruction of Kondo screening in Kondo lattice systems.

\section{Acknowledgements}
W.W thanks Karyn Le Hur and A.M.S. Tremblay for useful discussions. Z.L. and D.X.Y. are supported by NKRDPC-2017YFA0206203, NKRDPC-2018YFA0306001, NSFC-11974432, GBABRF-2019A1515011337, and Leading Talent Program of Guangdong Special Projects (201626003). W.W acknowledges the funding from the National Natural Science Foundation of China (Grant No. 41030053), and the Natural Science Foundation of Guangdong Province (Grant No. 42030030). This work was supported by a grant from the
Swiss National Supercomputing Centre (CSCS) under project ID S820.

\bibliographystyle{apsrev4-2}
\bibliography{heavyfermion2015}


\end{document}